# Channel Modeling of Human Somatosensory Nanonetwork: Body Discriminative Touch and Proprioception Perspective

Partha Pratim Ray

Dept. of Computer Sciences and Applications, Sikkim University, Gangtok-737102, India

parthapratimray@hotmail.com

*Abstract*- **Nanonetwork design and analysis has become a very interesting topic in recent years. Though this area of research is in its formative stage, it definitely posses a strong integrity in finding out numerous applications in medical and allied sciences. Nanonetworking is indeed a nature built foundation which comprises human intra body communications. Somatosensory system is the one of the critical and must have systems of human body. This literature concentrates on the body discriminative touch and proprioception mechanism of somatosensory system. This particular system is well architecture by medial lemniscal pathway, in human body for transduction of touch and proprioceptive information. This paper seeks out the novel communication channel model of somatosensory system. The working principle of the channel model is established by an equivalent Moore machine. A novel algorithm MLP is proposed after its name, medial lemniscal pathway. A novel naomachine and appropriate processing unit are also devised, based on the automaton.**

*Keywords*- **Nanonetwork, naomachine, somatosensory system, medial lemniscal pathway, moore machine, finite state automata, nanotechnlogy.**

## I. Introduction

Nanonetworking is a very fresh topic of interdisciplinary research. Actually nanonetworking is nano scale networking. It is a connected network between nanomachines, devices of nano or micrometer range which perform very simple tasks such as processing, storing data, sensing and corresponding actuation [1]. Nanonetworks are specialized in doing collaborative work comprising nanomachines extending its complexity and operation range [2]. Molecular and electromagnetic communication approaches have been proposed so far [3][4]. Being bio-inspired mechanism this paper assumed molecular communication as its basis. Molecular communications defines molecules as the communication building block while communicating from sender to receiver. Human body is a proper example of various interconnected nanonetworks such as cardiovascular, hormonal, neuronal, somatosensory, auditory, olfactory etc. communication between organs happen due to commands conveyed from Central Nervous Network (CNN) to Peripheral Nervous Network (PNN). CNN communicates the rest of human body as internet through the gateway of PNN. PNN is subdivided into two subnetworks such as, Somatic Nervous Subnetwork (SoNS) and Autonomic Nervous Subnetwork (AoNS). SoNS senses the 5 senses from different inputs and transmits the signal to CNN. While AoNS signals the motors and glands for actuation purpose conveyed by CNN.

The somatosensory systems coordinate the objects in our external environment through touch (i.e., physical contact with skin) and about the position and movement of our body parts (proprioception) through the stimulation of muscle and joints. The somatosensory systems also monitor the temperature of the body and provide information about painful, itchy and tickling stimuli. The sensory information processed by the somatosensory systems travels along different anatomical pathways depending on the information carried. For example, the posterior column-medial lemniscal pathway carries discriminative touch and proprioceptive information from the body, and this information is carried by main sensory trigeminal pathway from the face. Whereas, the crude touch, pain and temperature information is carried by spinothalamic pathways, and this same information is carried by the spinal trigeminal pathway from the face [5]. Somatosensory system actually governed by few modes specially, pain, temperature, touch and proprioception. Each has its own sub modality and corresponding sub sub modality. Each mode has its two types of pathways to communication with CNN, such as body and face.

Body discriminative touch and proprioception channel modeling has been taken as crucial to design its Moore machine and equivalent nanomachine. Though other communication pathways exist in the somatosensory system, a discriminative touch and proprioception pathway from body (medial lemniscal pathway) has been taken as a foundation of this paper.

Due to the advancement of Information Communication Technologies (ICT), human civilization has been reached to the top of the expectations. Researchers have performed very limited amount of work in nanonetworking field so far. Hence the need of improvising the same context arisen, which could be useful in terms of human health.Complex diseases procurement and medical diagnostic systems have been untouched, though. This literature assumed that such kind of difficulties could easily be sorted out if, similar artificial (rather say computational) model were devised. Such models might allow the medical diagnostic and medicinal therapies to work in different aspect that it had not been seen till now. Diseases arisen for communicational problems (i. e., neuronal malfunction) in human body could be prescribed instantly. Parallel advancement of nanotechnology would surely produce accessible nanomachines (theoretical aspect) in realty. Such machines indeed could run on the theoretical platform that would be presented here.

This paper has figured out the communication channel model behind the somatosensory system of body discriminative touch and proprioception (medial lemniscal pathways). A novel Moore machine model has been developed. An MLP algorithm (medial lemniscal pathway) is also devised. A novel nanomachine and its processor have also been presented in this paper.

This paper is organized as follows. Section II presents related work. Section III is about basics of somatosensory systems and pathways. Section IV explains communication channel modeling. Section V illustrates MLP algorithm. Section VI includes Moore machine model. Section VII presents nanomachine architecture. Section VIII concludes this paper.

## II. Related Work

Literature [6] proposes a Moore machine model, adapted by comprehending  biological interactions between nano scale neuro-spike communication and devises nano computer model based on Moore machine. Ray has recently proposed a novel automata modeling of hormonal molecular communication channel in human body which is followed by a nano machine design [7]. In another paper Ray has developed a Moore machine model of human auditory system. This paper also proposed a novel communication channel model of the same [8]. [9] discusses how the nervous system processes auditory information by combining knowledge from anatomy (to know which structures participate), electrophysiology (that tells us the properties of these structures), and finally, mathematical modeling (as a powerful tool for studying the mechanisms of the observed phenomena). Literature [10] presents various elementary models for significant intra-body molecular communication channels, including nanoscale neuro-spike communication channel, action potential-based cardiomyocyte molecular communication channel, and hormonal molecular communication channel along with multi-terminal nanonetworks extensions of channel models, with emphasis on the nervous, cardiovascular molecular, and endocrine nanonetworks. Thesis [11] represents the modeling of Finite State Automata (FSA) of Quorum Sensing mechanism in Bacteria and designs a nanomachine to implement Quorum Sensing. Paper [13] proposes synaptic Gaussian interference channel along with the characterization of power or firing rate of achievable rate region for the channel. [12] proposes an information theoretical model to understand the signaling mechanism of the molecular communication medium. In [14], an analytical framework that incorporates the effect of mobility into the performance of electrochemical communication among nanomachines is presented. Literature [15] proposes a stochastic dynamical model of noisy neural networks with complex architectures and discusses activation of neural networks by a stimulus, pacemakers and spontaneous activity.

## III. Basics of Somatosensory System

The somatosensory systems process information about, and represent, several modalities of somatic sensation (i. e., pain, temperature, touch, proprioception). Each of these modalities can be divided into sub-modalities, (e. g., pain into sharp, pricking, cutting pain; dull, burning pain; and deep aching pain etc.) [5]. This means that, sensation being stimulated by either naturally (e. g., by skin cooling) or artificially (e. g., by electrical stimulation of the neuron) is specified by the processed information of the neuron (i. e., cold skin). Similarly, cold somatosensory neuron does not respond to the heating of the skin. The modality specificity is determined by the somatosnsory receptors and its central connections which in turn form a somatosensory network.

Few kinds of stimuli are responsible for perception of stimulus in sensory systems. Tactile stimuli are external forces in physical contact with the skin that give rise to the sensations of touch, pressure, flutter, or vibration. Touch is obtained by involvement of minimal force on-or-by an object that produces very little distortion of the skin. A greater force that displaces the skin and underlying tissue arisen the pressure. Object movement or object flutter (20 to 50 Hz) or vibrations (100 to 300 Hz) are results of time varying tactile stimuli produce more complex sensations. Proprioceptive stimuli are arisen from internal forces that are generated by the position or movement of a body part. The dynamic changes in the forces applied to muscles, tendons and joints result the movement of a limb [5]. Proprioception is critical for maintaining posture and balance. Pain stimuli are the result of tissue-damaging sources of energy that may be external or internal to the body surface. Similarly, the sensation elicited on initial contact with the painful stimulus is sharp, cutting pain stimuli. A consequence of tissue inflammation follows the sensation of dull, burning pain [5] (Table 1).

Table 1. Sensory modalities of somatosensory system.

| Modality | Sub Modality | Sub-Sub Modality | Somatosensory Pathway (Body) | Somatosensory Pathway (Face) |
|---|---|---|---|---|
| Pain | sharp cutting pain | | Neospinothalamic | Spinal Trigeminal |
| | dull burning pain | | Paleospinothalamic | |
| | deep aching pain | | Archispinothalamic | |
| Temperature | warm/hot | | Paleospinothalamic | |
| | cool/cold | | Neospinothalamic | |
| Touch | itch/tickle, crude touch | | Paleospinothalamic | |
| | discriminative touch | touch | Medial Lemniscal | Main Sensory Trigeminal |
| | | pressure | | |
| | | flutter | | |
| | | vibration | | |
| Proprioception | Position: Static Forces | muscle length | | |
| | | muscle tension | | |
| | | joint pressure | | |
| | Movement: Dynamic Forces | muscle length | | |
| | | muscle tension | | |
| | | joint pressure | | |
| | | joint angle | | |

Peripheral organization of somatosensory systems has three components to describe. Peripheral somatosensory neurons: The cell bodies of the first-order (1°) somatosensory afferent neurons (pseudounipolar cells) are located in posterior root or cranial root ganglia (i. e., are part of the peripheral nervous system). A single process that divides to form a peripheral axon which travels to and ends in the skin, muscle, tendon or joint and a central axon which travels to and ends in the central nervous system is arisen by cell body. Somatosensory receptor organs: are located in specialized sensory receptor organs (e.g., the photoreceptors in the eye and the auditory and vestibular hair cells in the inner ear) or within a restricted part of the body (e.g., the taste buds in the mouth and the olfactory receptors in the olfactory mucosa of the nose). The somatosensory receptors and somatosensory stimuli are housed and delivered by these sensory receptors, to the receptors respectively. Sensory receptors: are specialized sensory receptor cells (e. g., the photoreceptors of the eye) are located in specialized receptor organs, produce receptor potentials, contain synaptic specializations, and release neural transmitters. Merkel cells are the only type of sensory receptor cell in the somatosensory system that can be found only in skin. Most peripheral axons of somatosensory 1° afferents also travel to skin, muscle or joint, branch near their terminal sites, and end in the skin, muscle, tendon or joint tissue. Somatosensory receptors are mainly of two types: cutaneous receptor and proprioceptive receptors. The subdivisions of these receptors are shown in Table 2 [5].

Table 2. Somatosensory receptors.

| Cutaneous Receptors | Type | Sensation | Signals | Adaptation |
|---|---|---|---|---|
| Meissner corpuscle | Encapsulated, layered | Touch: Flutter, Movement | Frequency/Velocity, Direction | Rapid |
| Pacinian corpuscle | Encapsulated, layered | Touch: Vibration | Frequency: 100-300 Hz | Rapid |
| Ruffini corpuscle | Encapsulated collagen | Touch: Skin Stretch | Direction, Force | Slow |
| Hair follicle | Unencapsulated | Touch: Movement | Direction, Velocity | Rapid |
| Merkel complex | Specialized epithelial cell | Touch, Pressure, Form | Location, agnitude | Slow |
| Free Nerve Ending | Unencapsulated | Pain, Touch, or Temperature | Tissue damage, Contact, Temperature change | Depends on information carried |
| Muscle Spindle | Encapsulated annulospiral and flower spray endings | Muscle stretch | Muscle length, velocity | Rapid initial transient and slow sustained |
| Muscle: Golgi Tendon Organ | Encapsulated collagen | Muscle tension | Muscle contraction | Slow |
| Joint: Pacinian | Encapsulated, layered | Joint Movement | Direction, velocity | Rapid |
| Joint: Ruffini | Encapsulated collagen | Joint pressure | Pressure, Angle | Slow |
| Joint: Golgi Organ | Encapsulated collagen | Joint torque | Twisting force | Slow |
| Muscle Spindle | Encapsulated annulospiral and flower spray endings | Muscle stretch | Muscle length, velocity | Rapid initial transient and slow sustained |

Medial lemniscal pathway: this pathway takes important role in carrying and processing of discriminative touch and proprioceptive information from the body (Figure 1). The information of discriminative touch and proprioception are kept separate from each other so as to map the signals at cerebral cortex efficiently. The peripheral first-order (1°) afferent and ending in the cerebral cortex

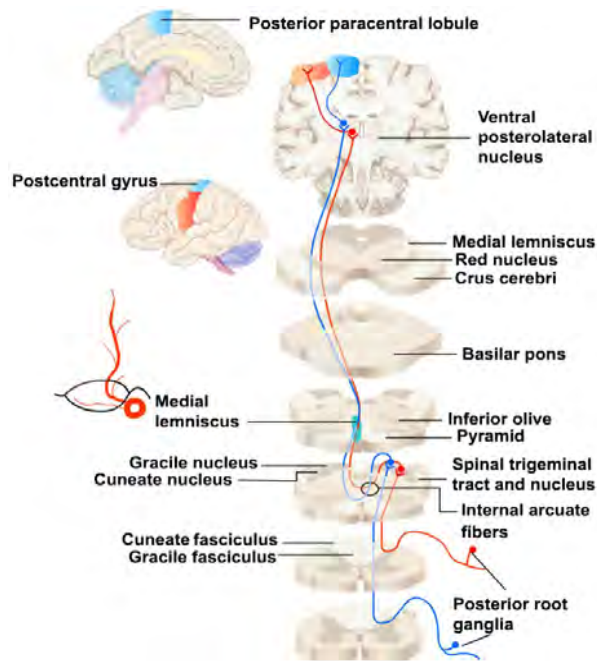

Figure 1. Medial lemniscal pathway.  (Courtesy: The University of Texas Health Science Center at Houston)

As per the explanation of Figure 1, The 1° medial lemniscal peripheral process starts at any of the somatosensory receptors such as skin, joints and muscles where Aβ axons are present, both in skin and joints and Group I and II types of axons are positioned in muscle. These axons of "coccygeal to mid-thorscic posterior roots" and "upper thoracic and cervical roots" ascend the spinal cord towards the ipsilateral gracile fasciculus and collect in the ipsilateral cuneate fasciculus, respectively. In medulla, 1° afferents in the gracile fasciculus synapse in gracile nucleus while cuneate fasciculus synapse in cuneate nucleus. The 2° medial lemniscal afferents carry cutaneous and proprioceptive information and get terminated in the core and surroundings of VPL, respectively. The axons of VPL 3° afferent neurons actually travel in the posterior limb of internal capsule and get terminated in the postcentral gyrus and postcentral lobule of the parietal lobe respectively. The postcentral gyrus and posterior paracentral lobule are known as primary somatosensory cortex which are also the primary cortical receiving areas of the somatosensory system.

## IV.  Channel Modeling of Medial Lemniscal Pathway

The communication channel model of medial lemniscal pathway is described in Figure 2. The communication system is divided into four segments, sensing, sender, communication channel and receiver. Proprioception and discriminative touch information- h(t) is sensed by the sensing segment (i.e., receptor endings). The information actually start journey from the sender (posterior root ganglia). The information is further processed by the communication channel (i.e., spinal cord and brain stem) after the addition with medial lemniscal afferent noise - m(t) , with feedback communication in it. This feedback generally occur due to the various neurotransmitter ejection associated to various lemniscal afferents (i. e., 1°, 2°, 3° etc.). The forwarded information is finally received by the receiver (i.e., primary somatosensory cortex). This communication pathway consists of posterior root ganglion, spinal nerve and brain stem. After posterior root ganglion, information passes through cuneate fasciculus and gracile fasciculus. Then it enters into spinal trigeminal tract and nucleus while passing by gracial and cuneate nucleus, both. Corresponding information is then reaches at medial lemniscus associated to inferior olive and pyramid in the surroundings. The further movement of information cuts through basilar pons and meets at red nucleus and crub cerebri region. At the last step of the information pathway, the information sends its journey at posterior paracentral lobule and postcentral gyrus, having surrounded  by ventral posterolateral nucleus.

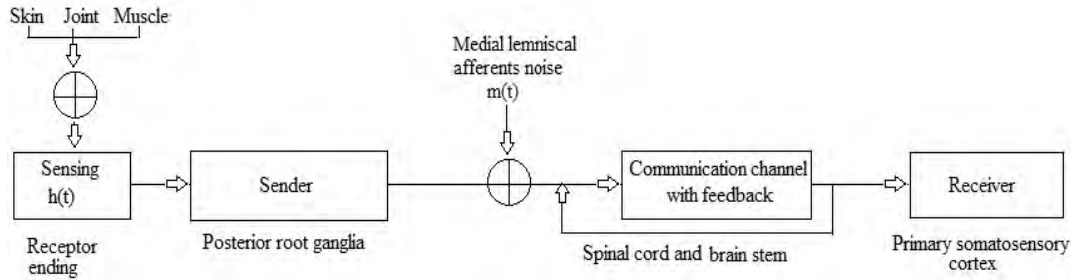

Figure 2. Communication channel model of medial lemniscal pathway.

## V. MLP Algorithm

This section provides a novel algorithmic approach towards the accumulation of Medial Lemniscal Pathway (MLP). The MLP algorithm (Figure 3) gives a pragmatic view of the somatosensory system with special attention to the discriminative touch and proprioception of human body. This algorithmic approach finds out the basic aspects of the medial lemniscal pathway communication mechanism. MLP is a linear algorithm which holds few parameters as sensory reception parameters, medial lemniscal afferent noise, $Na^+$ and $K^+$ density, neurotransmitter release factor etc. This algorithm definitely lays a foundation for ML pathway comprehension.

While analyzing the MLP algorithm it can be said that this algorithm clearly depicts very critical and crucial phases of information processing through the medial lemniscal pathway of somatosensory system. Particularly in five phases the MLP algorithm is divided. Sensing the signal and sourcing the information where cumulative signal reception- H(t) parameter is taken as the source of the communication path. In medulla, channeling factor S(t) is taken into consideration for information channeling. M(t) plays important role in 2° medial lemniscal afferent axons. Ventral posterolateral afferent noise- V(t) takes lead in 3° afferent neuron signal enhancement phase. At last, the signal reception at primary somatosensory cortex is measured by V(t). The most important portion of MLP algorithm is laid in between step 6 to 20, where a "while" loop is devised. Among S(t), M(t) and V(t), S(t) is the key of the whole system as this factor represents the information processing in the medulla of medial lemniscal pathway which coordinates the other parameters of MLP to proceed the information from the source to the destination. Hence, it can be said that the time complexity of MLP is $\mathbf{O}(M(t))$. Such a straight and linear algorithm has laid a preliminary foundation of medial lemniscal pathway to understand the body discriminative touch and proprioception. This novel attempt is indeed a remarkable achievement in nanonetworking research, done so far. This will surely help the followers of this field to move in a more charismatic stage of understanding the interactions between nanonetworks in human body.

## VI. Moore Machine Development

### A. Finite State Automata (FSA)

An Moore machine is a variant of finite state automaton which runs upon the present state. FSA is a mathematical model to conceive abstract machine which is a state at one time known as present state. When an event triggers up transition between states happened. Deterministic finite state automaton is a five-tuple as shown below. If A is a dfsa then it represents the below mentioned set. Deterministic FSA is described as below.

*A deterministic finite state automaton is a system A = {Q , ∑ , δ, $q_0$, F}, where Q is a finite set of the possible internal states of the automaton A, ∑ is a finite alphabet, $q_0$ is the initial state, δ is the transition function (δ: Q×∑→Q) and F is a subset of Q, the set of final or acceptance states.*

MLP: Medial Lemniscal Pathway Algorithm

*Begin*
*Inputs:*
$H_s(t)$: *signal reception at skin*
$H_j(t)$: *signal reception at joints*
$H_m(t)$: *signal reception at muscle*
$H(t)$: *cumulative signal reception at the receptors*
$H_{th}(t)$: *threshold value of cumulative signal reception*
$S(t)$: *medulla synapsing factor*
$S_{gf}(t)$: *synapsing of gracile fasciculus*
$S_{cf}(t)$: *synapsing of cuneate fasciculus*
$S_{th}(t)$: *threshold value of synapsing*
$A_{bs}(t)$: *ascending the brain stem*
$T_t(t)$: *terminating signal of thalamus in ventral posterolateral*
$T_c(t)$: *terminating cutaneous signal of thalamus in ventral posterolateral*
$T_p(t)$: *terminating  proprioception  signal*
$M(t)$: *medial lemniscal afferent noise*
$V(t)$: *ventral posterolateral afferent noise*
$V_{th}(t)$: *threshold value of ventral posterolateral afferent noise*
$V_v(t)$: *ventral posterolateral movement by potential enhancement*
$V_t(t)$: *ventral posterolateral termination potentialin parietal lobe*
$R(t)$: *receiving signal strength*
*Step1.     Sensing the signal and sourcing the information*
*Step2.     $H(t)= H_s(t)+H_j(t)+H_m(t)$*
*Step3.     If $(H(t))>H_{th}(t)$*
*Step4.             Then go to step 2*
*Step5.     Endif*
*Step6.     While (1)            {*
*Step7.             Channeling the information in medulla*
*Step8.             $S(t)= S_{gf}(t)+S_{cf}(t)$*
*Step9.             If $S(t)>S_{th}(t)$*
*Step10.             Then go to step 3*
*Step11.             Endif*
*Step12.     2° medial lemniscal afferent processing*
*Step13.     $M(t)= A_{bs}(t)+T_t(t)+T_c(t)+T_p(t)$*
*Step14.     If $M(t)>0$*
*Step15.             Then go to step 4*
*Step16.     Endif*
*Step17.     3° afferent neuron signal enhancement*
*Step18.     If $((V(t)= V_v(t)+ V_t(t))>V_{th}(t))$*
*Step19.             Then go to step 5*
*Step20.     Endif      }*
*Step21.  Signal receiving procurement*
*Step22.  If $((R(t)=H(t) + S(t)+ M(t) + V(t))!=0)$*
*Step23.         Then accept the signal*
*Step24.  Endif*
*End*

Figure 3.   Medial Lemniscal Pathway algorithm-MLP.

*B. Moore machine*

Moore machine is a variant of deterministic finite state automaton in which each state is bound to an output which depends upon present state only. It is a six-tuple set which resembles the dfsa but differs in output function and output alphabet. The model for the hormonal molecular communication channel is based on a Moore machine, since all the transitions are fixed, and an output is to be defined.

*A Moore machine is a six-tuple A = [Q, ∑, Λ, δ, τ, q₀], where Q is a finite set of the possible internal states of the automaton A, ∑ and Λ are finite alphabets for the input and the output, respectively, q₀ is the initial state, δ is the transition function (δ: Q×∑→Q) and τ is the output function (τ : Q→ Λ).*

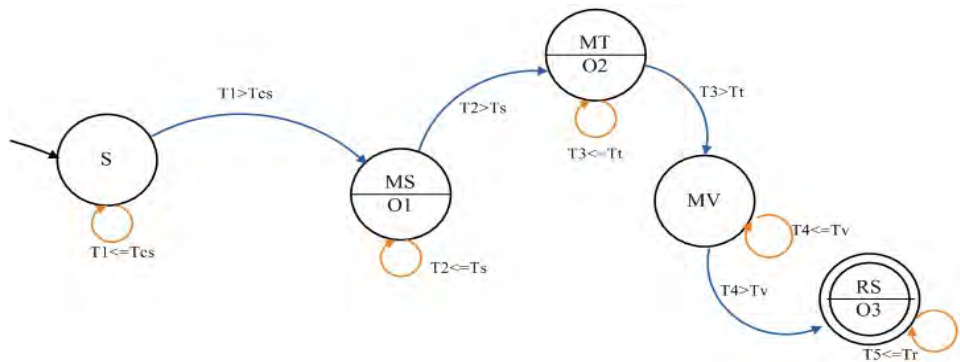

Figure 4.  Moore machine of medial lemniscal pathway mechanism.

Figure 4 presents the proposed Moore machine relevant to the medial lemniscal pathway communication mechanism. This state machine contains five different states. State S represents the start state of the machine. State MS illustrates the medulla synapse state. Later, MT defines medulla termination state. MV is another state which is ventral posterolateral state. Last, RS state, the receiving state.  The Moore machine presented in Figure 4 has five states:  Q={S, MS, MT, MV, RS}.

Table 3.  Input values and their assumptions.

| Input values | Assumptions |
|---|---|
| $T_1$ | Cumulative signal reception factor |
| $T_2$ | Synapsing factor |
| $T_3$ | Terminating factor |
| $T_4$ | Ventral posterolateral signal factor |
| $T_5$ | Reception factor |
| $T_{cs}$ | Cumulative signal reception threshold |
| $T_s$ | Synapsing factor threshold |
| $T_t$ | Terminating factor threshold |
| $T_v$ | Ventral posterolateral signal threshold |
| $T_r$ | Reception factor threshold |

Table 3 shows the various input values and their assumption, which have been taken as crucial step in Moore machine designing.  While, Table 4 presents the alphabets-∑ of the Moore machine. Alphabets ∑= {$d_1$, $d_2$, $d_3$, $d_4$, $d_5$, $d_6$, $d_7$, $d_8$, $d_9$, $d_{10}$}, are the primary basis of the state machine. The equivalent resultants are also pasted on the Table 4.  Table 5 depicts all the output alphabets Λ= {$O_1$, $O_2$, $O_3$}. The Moore machine start state is: $q_0$={S}. Final state of the Moore machine is F= {RS}. $d_x$/∈- presents the action/event concept where x-1,2,3,…,9.  ∈ is the event that means impossible or illegal event shown in Table 6.

Table 4. Alphabet-$\sum$.

| Input values | Resultants |
|---|---|
| $d_1$ | $T_1 <= T_{cs}$ |
| $d_2$ | $T_1 > T_{cs}$ |
| $d_3$ | $T_2 <= T_s$ |
| $d_4$ | $T_2 > T_s$ |
| $d_5$ | $T_3 <= T_t$ |
| $d_6$ | $T_3 > T_t$ |
| $d_7$ | $T_4 <= T_v$ |
| $d_8$ | $T_4 > T_v$ |
| $d_9$ | $T_5 <= T_r$ |

Table 5. Output- $\tau$.

| States | S | MS | MT | MV | RS |
|---|---|---|---|---|---|
| Output | $\epsilon$ | $O_1$ | $O_2$ | $\epsilon$ | $O_3$ |

Table 6. State transition table.

| | S | MS | MT | MV | RS |
|---|---|---|---|---|---|
| S | $d_1/ \epsilon$ | $d_2/ O_1$ | --- | --- | --- |
| MS | --- | $d_3/ O_1$ | $d_4/ \epsilon$ | --- | --- |
| MT | --- | --- | $d_5/ O_2$ | $d_6/\epsilon$ | --- |
| MV | --- | --- | --- | $d_7/ \epsilon$ | $d_8/O_3$ |
| RS | --- | --- | --- | --- | $d_9/ \epsilon$ |

## VII. Nanomachine design

A novel nanomachine architeture is presented in Figure 5. The nanomachine comprises of five units, namely input, storage, medial lemniscal processing, control and output unit. As the names suggest, the behavaioral characteristics are also done per their defination.

*Input:* This is the very basic and important unit of the machine. The starting phase of the medial lemniscal pathway is made through the cumulative signal (i. e., H(t)) reception at the receptors like skin, joints and muscles are which is here by correlated to this input unit.

*Storage Unit:* Sensed input signal is directed and stored temporarily in this storage unit. Further processing of the information might be carried out in this portion of the nanomachine. Basically simplest type of arithmetic and logical tasks (i. e., addition, subtraction, and, or and not) can be computed here. Data can be communicated between storage unit and control unit (see Figure 4). Feedback data can be obtained from output unit when there any signal is commanded by control unit.

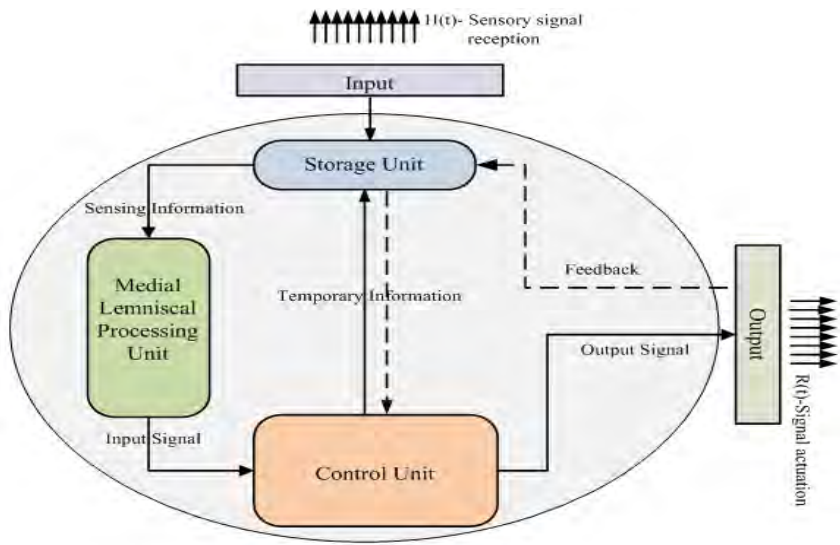

Figure 5. Nanomachine architecture.

*Medial Lemniscal Processing Unit:* Processing of the signal coming from input through storage unit is done at medial lemniscal processing unit. The Moore machine devised in earlier section is implemented in this unit. The organization of processing unit is presented in Figure 6. The unit contains clock signal, input signal (i. e., $T_1$, $T_2$, $T_3$, $T_4$, $T_5$, $T_{cs}$, $T_s$, $T_t$, $T_v$, $T_r$), output signal (i.e., $d_1$, $d_2$, $d_3$, $d_4$, $d_5$, $d_6$, $d_7$, $d_8$, $d_9$, $d_{10}$), and wait signal, computed as well as predefined signal inputs are processed by the intervention of control unit and produces outputs which are the alphabets of the Moore machine (Table 4). Along with this, wait signal is another component of the processing unit which might be useful for further computation of other connected computing units in the system periphery.

*Control Unit:* This unit of nanocomputer can be seen as the controller of the arithmetic, logical, and control signal, which are essential for proper functioning of nanomachine.

*Output Unit:* The overall processing of the nanomachine is finally turned into expected outcome value as signal which is further used for actuation or other bio-related computation.

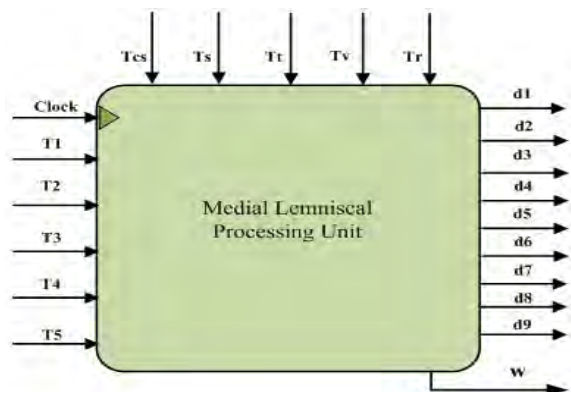

Figure 6. Processing Unit.

## VIII. Conclusion

A novel approach towards computational modeling of somatosensory system with special interest in discriminative touch and proprioception is proposed in this paper. Medial lemniscal pathway has been the target area on which an MLP algorithm has been proposed having complexity of $\mathbf{O}(M(t))$. A novel communication channel model is devised in this paper. A novel Moore machine and corresponding architecture and organization of nanomachine and its processing unit have been developed. Lack of theory and mathematical aspects in this field, is one of its important limitations of this paper. Simulation environment should also be developed to imply this practicality. Though this research is in its infant stage, the novel mechanisms regarding nanonetworking would surely pave the paths of innovative and advanced research in the area of nanonetworking research.

# References

[1]  Akyildiz I F, Brunetti F, Blazquez C, Nanonetworks: A New Communication Paradigm. Computer Networks Elsevier Journal. 52(12): 2260-2279, 2008.

[2]  Nanonetwork, http://en.wikipedia.org/wiki/ Nanonetwork, 2012.

[3]  Moore M, Enomoto A, Nakano T, Egashira R, Suda T, Kayasuga A, Kojima H, Sakakibara H, Oiwa K,  A Design of a Molecular Communication System for Nanomachines Using Molecular Motors. Proceeding of Fourth Annual IEEE Conference on Pervasive Computing and Communications and Workshops, 2006.

[4]  Akyildiz I F, Jornet J M,  Electromagnetic Wireless Nanosensor Networks. Nano Communication Networks Elsevier Journal. 1(1): 3-19, 2010.

[5]  Dougherty P, Somatosensory systems. http://neuroscience.uth.tmc.ed/s2/chapter02.html, 2012.

[6]  Ray P P, Nano Computer Design Based on Intra Body Nanoscale Neuro-Spike Communication: a Nanonetwork Paradigm. accepted in  International Conference on Communications, Devices and Intelligent Systems, 2012.

[7]  Ray P P, Automata Modeling of Hormonal Molecular Communication Channel in Human Body. accepted in  International Joint Journal Conference in Computer Electronics and Electrical, 2012.

[8]  Ray P P, Communication Channel Modeling and Automata Designing of Human Auditory System. accepted in  International Joint Journal Conference in Computer Electronics and Electrical, 2012.

[9]  Borisyuk A, Physiology and mathematical modeling of the auditory system. Lecture Notes in Mathematics. 1860: 107-168.

[10]  Malak D, Akan O B, Molecular communication nanonetworks inside human body. Nano Communication Network.3:19–35, 2012.

[11]  Cavalle S A, Automata Modeling of Quorum Sensing for Nanocommunication Networks. Masters Thesis, Universitat Polit_ecnica de Catalunya, Barcelona, 2010.

[12]  Liu J Q, Nakano T, An information theoretic model of molecular communication based on cellular signaling, in Bio-Inspired Models of Network, Information and Computing Systems, pp. 316-321, 2007.

[13]  Balevi E, Akan O B, Synaptic Gaussian multiple access channel. submitted for publication, 2011.

[14]  Guney A, Atakan B, Akan O B, Mobile Ad Hoc Nanonetworks with Collision-based Molecular Communication.  IEEE Transactions on Mobile Computing, 11(3):353- 366 , 2012.

[15]  Goltsev A V, Abreu F V de, Dorogovtsev S N, Mendes J F F, Stochastic cellular automata model of neural networks. arXiv:0904.2189v3:1-10, 2010.